\documentstyle[psfrag,preprint,graphicx,tighten,floats,aps,amsmath,amsfonts,amssymb]{revtex}

\def\H{{\cal H}}

\def\u(1){{\rm u(1)}}

\def\lp{{\ell}_{\rm Pl}}
\begin{document}

%\date\today

\title{Black hole entropy from Quantum Geometry}
\author{Marcin Domagala$^{1}$ and  Jerzy Lewandowski${}^{1,2,3}$}
\address{1. Instytut Fizyki Teoretycznej,
Uniwersytet Warszawski, ul. Ho\.{z}a 69, 00-681 Warszawa, Poland\\ 2.
Physics Department, 104 Davey, Penn State, University
Park, PA 16802, USA\\ 3.
Max Planck Institut f\"ur Gravitationsphysik, Albert Einstein
Institut, 14476 Golm, Germany}

\maketitle

\begin{abstract}
Quantum Geometry (the modern Loop Quantum Gravity using graphs and 
spin-networks instead of the loops) provides microscopic degrees of freedom 
that account for the black-hole entropy. However, the procedure for
state counting used in the literature contains an error and the
number of the relevant horizon states is underestimated. In our
paper a correct method of counting is presented.  Our results lead
to a revision of the literature of the subject. It turns out that
the contribution of spins greater then 1/2 to the entropy is not
negligible. Hence, the value of the Barbero-Immirzi parameter
involved in the spectra of all the geometric and physical
operators in this theory is different than previously derived.
Also, the conjectured relation between Quantum Geometry and the
black hole quasi-normal modes should be understood again.
\end{abstract}

%\maketitle

%\maketitle %\pacs{04.60Pp, 04.60.Ds, 04.60.Nc, 03.65.Sq}
\bigskip
\indent\emph{Pacs {04.60Pp, 04.60.Ds, 04.60.Nc, 03.65.Sq}}

\vfill\break
\section{Introduction and results}
Quantum Geometry
\cite{thiembook,rovbook,alrev}  provides microscopic degrees of freedom
that account for the black-hole entropy. The issue was raised in  works
by Krasnov \cite{krasentropy} and  Rovelli
\cite{roventropy}. The final quantum Isolated Horizon framework was
introduced by  Ashtekar, Baez, Corichi and Krasnov \cite{abk}. The framework describes
a black-hole in equilibrium
surrounded by a quite arbitrary space-time. In the canonical formulation of the theory,
the classical phase space is restricted to the set of  space-times containing
a black-hole of a fixed area $a$.  After quantization, the kinematic Hilbert space
is the tensor product of the horizon Hilbert space
$\H_{S}$ and the bulk Hilbert space  $\H_{\Sigma}$. Elements of $\H_{S}$
are identified with the black-hole kinematic quantum states. They
represent quantum excitations of the $U(1)$ spin connection defined on the horizon.
Mathematically, they are described by the quantum  Chern-Simons theories
on a punctured sphere, where all possible sets of punctures are admitted.
Therefore, they are labeled by certain sequences of numbers (referred to
as spins) assigned to the punctures.
The bulk states, that is elements of  $\H_{\Sigma}$,  on the other hand,
are described by Quantum Geometry, and
define the quantum area of the horizon \cite{rsarea,alarea}.
The quantum constraints commute with the quantum horizon area  operator. Most
importantly, all the solutions whose quantum areas fall into any given finite
interval $[a -\delta a, a +\delta a]$ can be labeled by a finite number of the
black-hole states (and by other labels corresponding to the bulk states).
That number defines the Ashtekar-Baez-Corichi-Krasnov black-hole entropy. However, the procedure
introduced in \cite{abk} for state counting contains a spurious constraint on
admissible sequences of the labels, and the number of the relevant horizon states
is underestimated. The goal of our work\footnote{This project was a part of the
diploma thesis of one of the authors \cite{pinek}.} is to point out this error,
to formulate a correct  counting procedure, and to study the consequences.
Several  conclusions of \cite{abk} have to be re-examined:
\begin{itemize}
\item[$(i)$] the statement that the leading contribution to the entropy comes entirely from
the simplest states (characterized by punctures  with labels $\pm 1/2$),
\item[$(ii)$] the confirmation of the Bekenstein and Hawking formula \cite{bekhaw},
\item[$(iii)$] the determination of  the value of the  free parameter $\gamma$ introduced in Quantum Geometry  by
Barbero and Immirzi,  as
\begin{equation}
\gamma_{\rm ABCK} = \frac{\ln 2}{\pi\sqrt{3}}.\label{gammaabk}
\end{equation}
\end{itemize}
%Indeed, we will see, that the first one and and last one are not true.
The Barbero-Immirzi parameter labeles different sectors of Quantum Geometry.
The spectra of all the geometric and physical operators depend on $\gamma$.
For example the quantized area is proportional to $\gamma$.

Therefore, in our paper we go back to the issue of calculating the entropy within
the Quantum Geometry and the Ashtekar-Baez-Corichi-Krasnov framework.  We formulate an
exact combinatoric problem whose solution is the entropy: we carefully define a set
of sequences freely labeling the black-hole quantum states which correspond to
eigenvalues of the quantum area operator equal or less than $a$.
The entropy is  $\ln N(a)$, where $N(a)$ is the number  of elements of
this set.  Next, we address the issue of the proportionality of the entropy to
the area. We consider the ratio
\begin{equation}
\frac{\ln N(a)}{a}
\end{equation}
which is expected to be constant in the limit as $a\rightarrow \infty$.
We find that for  large $a$,
\begin{equation}
\frac{\ln 2}{4\pi\gamma\lp^2}\, a\
\le\ \ln N(a) \le\ \frac{\ln 3}{4\pi\gamma\lp^2}\, a\, . \label{entropyest}
\end{equation}
As before (in \cite{abk}), the Barbero-Immirzi parameter $\gamma$ can be fixed
by assuming the agreement with the Bekenstein-Hawking entropy
(the proportionality of the entropy to the area in the leading
order is  not proved yet, at this point, but see below).
Then, (\ref{entropyest})  provides  lower and upper bounds for $\gamma$,
\begin{equation}
\frac{\ln 2}{\pi} \le \gamma \le \frac{\ln 3}{\pi}\label{gammaineq}.
\end{equation}
These results lead to a  revision of the literature of the
subject.

To begin with, they show that  the number of states characterized
by the lowest spin labels  is essentially smaller than the number
of all the states contributing to a given value of the horizon
area. Therefore, the horizon states labeled by higher spins do
contribute to the leading term of the entropy.  In the
consequence, the ABCK model of quantum black hole  does not agree
with Wheeler's heuristic model `It from bit'. The (contradicted
here) conclusion $(i)$ listed above, had been used in \cite{Dre}
to point out a remarkable relation between the black hole
quasi-normal modes, the Bekenshtein-Hawking entropy and the ABCK
entropy provided  the minimal spin of the quantum excitations of
the horizon is $1$ rather then $1/2$. In view of our correction,
the meaning of that observation of \cite{Dre} should be understood
again.

Next, the conclusion $(iii)$ listed above is not true: the value
of the parameter $\gamma$ can not be that of (\ref{gammaabk}).

We were not able to verify the proportionality of the horizon
entropy to the area ourselves.  However, our combinatoric
formulation of the problem of calculating the entropy derived in
this paper has been later solved in an exact way  in  the
accompanying paper \cite{meis}. The number $\ln N(a)$ has been
calculated rigorously therein, and the result is
\begin{equation}
\ln N(a)\ =\ \frac{\gamma_{\rm M}}{\gamma}\frac{a}{4\lp^2} - \frac{1}{2}\ln a
+  O(1)
\end{equation}
in the limit of large horizon area $a$, where the value of the
parameter $\gamma_{\rm M}$ can be calculated at arbitrary
accuracy. In conclusion, taking into account the results of
\cite{meis},  the proportionality of the entropy to the area in
the leading order for large $a$ is confirmed, and even the
sub-leading terms are calculated.

In the last section we derive the probability distribution of the
spin at a given puncture. The physical meaning of these new
results will be discussed in a joint paper \cite{adlm}.

\section{The horizon quantum states}
Our departure point  is the quantum Isolated Horizon theory
introduced in \cite{abk}. A classical (weakly) isolated horizon is
a null, non-expanding world-surface  of a space-like 2-sphere $S$,
equipped with a null vector field of a constant self-acceleration.
This idea gave rise to a new quasi-local theory of  black-hole  in
equilibrium interesting also from purely classical point of view
\cite{IH}.  The local degrees of freedom of space-time geometry
out of  the horizon are not restricted. An additional assumption
satisfied by the isolated horizon geometry considered in
\cite{abk} is the spherical symmetry (see \cite{alrev} for a
discussion of the general case). The classical phase space
introduced in the Ashtekar-Baez-Corichi-Krasnov framework consists
of all the gravitational fields which can exist in a neighborhood
of an isolated horizon. The location and area $a$ of the horizon
are fixed. In the corresponding $3+1$ Hamiltonian framework the
horizon is represented by a 2-sphere $S$, whereas the remaining
part of space under consideration (bulk) is the exterior region, a
3-manifold $\Sigma$ bounded by $S$. The only kinematic
 degrees of freedom of the horizon $S$  are given by a U(1) connection defined on a
spin bundle over $S$.
The gravitational field data is defined on $\Sigma$. The bulk and
the horizon data are subject to a consistency condition ensuring that the world-surface
of $S$ is indeed an isolated horizon. In the quantum theory of this system,
the bulk gravitational field is described by  Quantum Geometry. The
quantum consistency condition between the bulk and the horizon implies that
the quantum degrees of freedom of the horizon should be described by a union
of the Chern-Simons theories on arbitrarily punctured $S$.

Below, we outline the elements of the quantum theory which play a role in the entropy
calculation.
The kinematic Hilbert space of the quantum Isolated Horizon theory is
{\it contained} in the tensor product   ${\cal H}_S\otimes{\cal H}_{\Sigma}$, where
 ${\cal H}_S$ is the Hilbert space of the quantum horizon degrees of freedom
whereas ${\cal H}_{\Sigma}$ is the Hilbert space of the quantum geometry
defined in the bulk $\Sigma$.

To  characterize  the bulk quantum geometry  Hilbert space ${\cal H}_{\Sigma}$,
it is convenient to use a certain orthogonal decomposition adapted to the boundary $S$.
Consider a finite set of points,
\begin{equation}
{\cal P}\ = \{p_1,...,p_n\}\subset S,
\end{equation}
and a labeling of elements of ${\cal P}$ by numbers $j=(j_1,...,j_n)$, and $m=m_1,...,m_n)$
where
\begin{equation}
j_i\in \frac{1}{2}{\mathbb{N}},\ \ \ \
m_i\in \{-j_1,...,j_i\}\label{jm}
\end{equation}
for $i=1,...,n$. The $j_i$ labels are assumed not to vanish,
\begin{equation}
j_i\not=0\, .\label{jnot0}
\end{equation}
The space ${\cal H}_{\Sigma}$  is  the orthogonal sum
\begin{equation}
{\cal H}_{\Sigma}\ =\ \bigoplus_{({\cal P},j,m)}{\cal H}_{\Sigma}^{{\cal P},j,m},\label{bulkdec}
\end{equation}
where ${\cal P}$ runs through all the finite subsets of $S$,
$(j,m)$ through all the finite labeling  (\ref{jm},\ref{jnot0}),
and the empty set ${\cal P}=\emptyset$ is also associated a
(non-trivial) single Hilbert space ${\cal
H}_{\Sigma}^{\emptyset}$. Each of the subspaces ${\cal
H}_{\Sigma}^{{\cal P},j,m}$ is a certain (infinitely dimensional)
Hilbert space. To understand the (quantum) geometric  meaning of
these subspaces, consider any piece $S'$ of the horizon $S$ (more
precisely, $S'$ is an arbitrary 2-sub-manifold in the 2-sphere
$S$). The quantum area operator $\hat{A}_{S'}$ is  defined  in
${\cal H}_{\Sigma}$. Another operator  of the quantum geometry in
$\Sigma$ assigned to $S'$ is the flux  $\hat{E}_{S'}$ of the
vector field normal to $S$.\footnote{ There are two subtleties,
which may be intriguing for the reader. The first one is that
classically the flux of a vector field normal to a surface {is}
the same as the area of the surface. In our quantum case however,
the vector field is a quantum operator itself, in fact it is one
of the SU(2) spin operators. The area, on the other hand, is given
by the total SU(2) spin operator. The second subtlety is, that
$S'$ contained in the {\it horizon} is assigned the area and flux
by the {\it bulk} geometry.} Every subspace  ${\cal
H}_{\Sigma}^{{\cal P},j,m}$ is the eigenspace of the operators.
The corresponding eigenvalues are:
\begin{eqnarray}
a_{S'}^{{\cal P},j}\ &=&\ 8\pi\gamma\lp^2\sum_{p_i\in {\cal P}\cap S'}\sqrt{j_i(j_i+1)},\\
e_{S'}^{{\cal P},m}\ &=&\ 8\pi\gamma\lp^2\sum_{p_i\in {\cal P}\cap S'}m_i
\end{eqnarray}
where only those points $p_i\in {\cal P}$ contribute which are contained also in $S'$,
and $\gamma>0$ is a free parameter of Quantum Geometry known as
Barbero-Immirzi parameter.
In particular, if we take for  $S'$  the whole horizon $S$, then the corresponding
eigenvalue of the horizon area operator is (we drop ${\cal P}$ because only
the set of the values of $j$ matters)
\begin{equation}
a_{S}^{j}\ =\  8\pi\gamma\lp^2\sum_{i}\sqrt{j_i(j_i+1)}.\label{area}
\end{equation}
Of course the quantum area and flux  operators are defined also for  2-dimensional
sub-manifolds of $\Sigma$ which do not intersect $S$, and there are also
other interesting operators of Quantum Geometry acting  in  ${\cal H}_{\Sigma}$
but they are not used in the Isolated Horizon framework.

We turn now to the horizon Hilbert space ${\cal H}_{S}$. Recall, that
the ABCK framework requires fixing an arbitrary value for the classical horizon area $a$.
In order to quantize consistently the horizon degrees of freedom,
it is assumed in \cite{abk} that the fixed classical area $a$ is quantized in the
following way,
\begin{equation}
a\ =\ 4\pi\gamma \lp^2 k,\ \ \ k\in {\mathbb{N}},
\end{equation}
where $k$ is arbitrary. The horizon Hilbert space can also be orthogonally decomposed
in a way analogous to (\ref{bulkdec}). In this case the labeling set
consists of all the finite sequences $\vec{\cal P}$ of points in $S$, labeled by
arbitrary non-zero integers defined modulo $k$, and whose sum is zero modulo $k$,
\begin{eqnarray}
\vec{\cal P}\ &=&\ (p_1,...,p_n)\, \ \ \ b\ =\ (b_1,...,b_n)\, ,\\
b_i\ &\in&\ {\mathbb{Z}}_k,\ \ \ i=1,...,n, \ \ \ \sum_{i=1}^nb_i=0\, .\label{phi}
\end{eqnarray}
To every  labeled sequence $(\vec{\cal P},b)$ there is assigned a 1-dimensional  Hilbert space ${\cal H}_S^{\vec{\cal P},b}$. The points $p_i$ are often called
punctures, and for every $p_i$, the  $U(1)$ element $e^{i\frac{2\pi b_i}{k}}$ is the
eigenvalue
of the holonomy transport along a circle containing $p_i$. The full horizon Hilbert space
can be written as the orthogonal sum
\begin{equation}
{\cal H}_S = \bigoplus_{(\vec{\cal P},b)}{\cal H}_S^{\vec{\cal P},b},
\end{equation}
where to the labeling set of pairs $(\vec{\cal P},b)$  we
add the empty sequence which labels a single, 1-dimensional space
${\cal H}_S^{\vec{\emptyset}}$.

Now, the quantum condition that $S$ be a section of a spherically symmetric isolated
horizon is introduced in the product Hilbert space
\begin{equation}
{\cal H}_S\otimes{\cal H}_\Sigma\ =\
\bigoplus_{(\vec{\cal P},b)}\bigoplus_{({\cal P}',j,m)} \left({\cal H}_S^{\vec{\cal P},b}
\otimes{\cal H}_{\Sigma}^{{\cal P}',j,m}\right).
\end{equation}
It involves the holonomy operators acting in ${\cal H}_S^{\vec{\cal P},b}$ on the
one hand and the  flux operators acting in  ${\cal H}_{\Sigma}^{{\cal P}',j,m}$
on the other hand (see \cite{abk} or \cite{alrev} for a summary).
The set of solutions is the orthogonal sum of products
${\cal H}_S^{\vec{\cal P},b}\otimes{\cal H}_{\Sigma}^{{\cal P},j,m}$
such that the points coincide modulo the ordering
\begin{equation}
\vec{\cal P}=(p_1,...,p_n), \ \ \ \ {\cal P}\ =\ \{p_1,...,p_n\}\label{c1}
\end{equation}
and the labelings agree with each other in the following way
\begin{equation}
b_i = -2 m_i\ \ {\rm mod}\ \  k.\label{c2}
\end{equation}
In conclusion, the kinematic Hilbert space of states in the ABCK
quantum Isolated Horizon framework is the orthogonal sum with
respect to all the labeled sequences of points in $S$ which
satisfy (\ref{c1},\ref{c2}),
\begin{equation}
{\cal H}_{\rm kin}\ =\ \bigoplus_{\vec{\cal P},j,m}
{\cal H}_S^{\vec{\cal P},b(m)}\otimes{\cal H}_{\Sigma}^{{\cal P},j,m}\, ,\label{kin}
\end{equation}
where we emphasized the dependence of each sequence $b=(b_1,...,b_n)$
 on
a sequence  $m=(m_1,...,m_n)$  via (\ref{c2}).

The final step of the ABCK framework is to take into account the quantum
Einstein constraints \cite{thiembook,rovbook,alrev} adjusted to the
Isolated Horizon framework \cite{abk}. The  vector constraints
generate diffeomorphisms of $\Sigma$ which preserve the boundary $S$.
The diffeomorphisms act naturally in  ${\cal H}_{\rm kin}$ and this action
is unitary. In the Quantum Geometry framework, the quantum vector constraints come down
to the diffeomorphism invariance.
Whereas there is only one normalizable diffeomorphism invariant state
in ${\cal H}_{\rm kin}$, the invariance condition is imposed  in a certain
larger space (dual to an appropriate subspace of  ${\cal H}_{\rm kin}$).
A Hilbert space of solutions is constructed via the diffeomorphism averaging
procedure whose Quantum Geometry version is well described in
\cite{thiembook,alrev}. It is particularly easy to explain the result of the
analogous
diffeomorphism averaging suitable for the  space ${\cal H}_S$ itself. Assign to every
finite sequence $b=(b_1,...,b_n)$ defined in (\ref{phi}) a 1-dimensional
Hilbert space ${\cal H}_S^{b}$, and define
\begin{equation}
{\cal H}_{S,{\rm phys}}\ =\ \bigoplus_{b}  {\cal H}_S^{b}\, .\label{Sphys}
\end{equation}
The averaging applied to the full space ${\cal H}_{\rm kin}$ provides a Hilbert
space of the following structure
\begin{equation}
{\cal H}\ =\ \bigoplus_{j,m}\tilde{{\cal H}}^{b(m),j,m},\label{tphys}
\end{equation}
where $b=(b_1,...,b_n)$, $j=(j_1,...j_n)$ and
$m=(m_1,...,m_n)$ are arbitrary triples of finite
sequences such that (\ref{jm},\ref{phi},\ref{c2}) holds.
The remaining constraints are the scalar and the Gauss constraint.
The Gauss constraint amounting to the invariance with respect to
the local SU(2) rotations of the spin frames is already satisfied
at $S$ due to (\ref{c2}). The scalar constraint, on the other hand,
has been solved at the horizon already on the classical level.
 Since the quantum numbers $(j,m)$ used in (\ref{phys}) characterize
the quantum geometry of bulk $\Sigma$ right at the boundary $S$, the
plausible, but non-trivial assumption made in \cite{abk}
is that the bulk scalar constraints with laps functions vanishing
on $S$ do not affect $(j,m)$, and that for every choice of $(j,m)$
there is a solution of the bulk scalar constraints.

We are lead to a conclusion, that the final physical Hilbert
${\cal H}_{\rm phys}$ of the quantum states can be characterized as
\begin{equation}
{\cal H}_{\rm phys}\ =\ \bigoplus_{j,m}
{\cal H}^{b(m),j,m},\label{phys}
\end{equation}
where $b=(b_1,...,b_n)$, $j=(j_1,...j_n)$ and
$m=(m_1,...,m_n)$ are arbitrary triples of finite
sequences such that (\ref{jm},\ref{phi},\ref{c2}) is satisfied,
and none of the spaces ${\cal H}^{b(m),j,m}$
is empty. Since the horizon area operator $\hat{A}_S$  commutes with all
the constraints in this framework, it passes to the physical
Hilbert space. It also follows from the details of the
diffeomorphism averaging map, that each of the subspaces
 ${\cal H}^{b(m),j,m}$ is an eigenspace corresponding
to the eigenvalue (\ref{area}). In this way, {\it the
sequences  $j$ are responsible for the  area assigned to the
2-surface $S$ of the horizon by the bulk Quantum Geometry,  whereas
the sequences $b$ represent the intrinsic quantum degrees
of freedom of the horizon.} This is the key conclusion of
the ABCK framework.

\section{The combinatoric formulation of the problem}
From now our treatment starts to differ from that of \cite{abk}.
Our aim is to define the  number of the quantum states
of the horizon which  contribute  to the entropy in a form of a
clear and precise combinatoric formula.

Recall that the classical area $a$ of the horizon is fixed
as
\begin{equation}
a\ =\ 4\pi\gamma\lp^2 k,\ \ \ k\in {\mathbb{N}}.
\end{equation}
Calculation of the horizon entropy consists in counting those quantum horizon
states, elements of the Hilbert space ${\cal H}_{S,{\rm phys}}$ (\ref{Sphys})
labeled by the finite sequences defined in (\ref{phi})  (the number $n$ of
the entries can be arbitrary) which  correspond to
subspaces ${\cal H}^{b(m),j,m}$ in (\ref{phys}) such that
the area operator eigenvalue $a_S^j$ satisfies the following
inequality\footnote{Assuming just inequality
instead of considering an interval $[a-\delta a,a+\delta a]$ simplifies
the calculation. Knowing the result for all $a$ one can always consider the
interval. The result, at least in our case, is the same.}
\begin{equation}
a_S^j\ =\ 8\pi\gamma\lp^2\sum_{i=1}^n\sqrt{j_i(j_i+1)}\ \le\ a.\label{main}
\end{equation}
Combinatorially, the task amounts to counting  the finite sequences
$(b_1,...,b_n)$  of elements of ${\mathbb{Z}}_k$,
such that the following two conditions $(i)$ and $(ii)$ are satisfied:
\begin{equation}
(i)\ \ \ \ \ \ \ \ \ \ \ \ \ \ \ \ \ \ \ \ \ \ \ \ \ \sum_{i=1}^nb_i=0\, , \label{phi2}
\end{equation}
$(ii)$ there exist two sequences: a sequence $(j_1,...,j_n)$ consisting of
non-zero elements of $\frac{1}{2} {\mathbb{N}}$ which satisfies the inequality
(\ref{main}), and  a sequence $(m_1,...,m_n)$ such that
\begin{eqnarray}
b_i &=& -2 m_i\ \ {\rm mod}\ \  k\, ,\ \label{phim}\\
\ \ {\rm and}\ \ m_i\ &\in&\ \{-j_i,-j_i+1,...,j_i\}\, ,\label{mj}
\end{eqnarray}
for every $i=1,,,n$.

This recipe can be  simplified.  It is easy to
eliminate the $j_i$'s by noting that, given a sequence
$(m_1,...,m_n)$, there is a sequence $(j_1,...,j_n)$
such that (\ref{mj}) and (\ref{main}) if and only if
\begin{equation}
8\pi\gamma\lp^2\sum_{i=1}^n\sqrt{|m_i|(|m_i|+1)}\ \le\ a.
\end{equation}
Next, it follows from (\ref{phi2}, \ref{phim}) that
\begin{equation}
\sum_{i=1}^n m_i\ =\ n'\frac{k}{2}\, ,
\end{equation}
with some $n'\in {\mathbb{Z}}$. But then,
\begin{eqnarray}
a\ =\ 4\pi\gamma\lp^2k\ &\ge&\ 8\pi\gamma\lp^2\sum_{i=1}^n
\sqrt{|m_i|(|m_i|+1)}>\\
   &>& 8\pi\gamma\lp^2\sum_{i=1}^n|m_i|\ \ge\
8\pi\gamma\lp^2|\sum_{i=1}^nm_i|\ =\ \\
\ &=&\ 4\pi\gamma\lp^2k|n'|.
\end{eqnarray}
Hence
\begin{equation}
\sum_{i=1}^nm_i=0, \ \ {\rm and\ \ also}\ \ \ \sum_{i=1}^n|m_i|\le\frac{k}{2}\, .
\label{mzero}
\end{equation}
Finally, notice that given a sequence $(b_1,...,b_n)$ of elements
of ${\mathbb{Z}}_k$ there is exactly one sequence $(m_1,...,m_n)$
of elements of $\frac{1}{2}{\mathbb{Z}}$ such that
(\ref{phim}) and (\ref{mzero}) holds. In this way, we have derived the following
formula for the entropy counting:
\bigskip

{\it The entropy $S$ of a quantum horizon of the classical area $a$ according
to Quantum Geometry  and the Ashtekar-Baez-Corichi-Krasnov \cite{abk} framework
is
\begin{equation}
S \ =\ \ln N(a),
\end{equation}
where $N(a)$ is $1$ plus the number of all the finite sequences $(m_1,....,m_n)$
of non-zero elements of $\frac{1}{2}{\mathbb{Z}}$, such that the following
equality and inequality are satisfied:
\begin{equation}
\sum_{i=1}^nm_i\ =\ 0,\ \ \ \ \sum_{i=1}^n\sqrt{|m_i|(|m_i|+1)}\ \le\
\frac{a}{8\pi\gamma\lp^2},\label{N(a)}
\end{equation}
where $\gamma$ is the Barbero-Immirzi parameter of Quantum Geometry.}
\bigskip

The extra term $1$ above comes from the trivial sequence.

\section{The entropy calculations}
To find an upper bound for the number $N(a)$  introduced in the previous
section (recall that $a=4\pi\gamma\lp^2k$, and $k\in {\mathbb{N}}$), consider the set $M^+_k$ of all the finite sequences
$(m_1,...,m_n)$ of non-zero elements of $\frac{1}{2}{\mathbb{Z}}$
which satisfy the inequality in (\ref{mzero}), but do not necessarily sum
to zero,that is,
\begin{equation}
M^+_k\ :=\ \{ (m_1,...,m_n)\ \Big|\ n\in{\mathbb{N}},\
0\not=m_1,...,m_n\in\frac{1}{2}{\mathbb{Z}},\
\sum_{i=1}^n |m_i|\ \le\ \frac{k}{2}\}\,.
\end{equation}
Let $N^+_k$ be the number of elements of $M^+_k$ plus $1$ (the empty sequence).
Certainly,
\begin{equation}
N(a)\ \le\ N^+_k
\end{equation}
Next, since $k$ is arbitrarily fixed integer, let it become  a variable
of the sequence $(N^+_0,N^+_1,...,N^+_k,...)$. To establish a recurrence
relation satisfied by the sequence $(N^+_0,N^+_1,...,N^+_k,...)$, notice
that if $(m_1,...,m_n)\in M^+_{k-1}$, then both $(m_1,...,m_n,\frac{1}{2}),\
(m_1,...,m_n,
-\frac{1}{2})\in M^+_{k}$. In the same way, for arbitrary natural $0<l\le k$,
\begin{equation}
(m_1,...,m_n)\in M^+_{k-l}\ \Rightarrow\ (m_1,...,m_n,\pm\frac{1}{2}l)
\in M^+_k.
\end{equation}
Obviously, if we consider all $0<l\le k$, and all the sequences
$(m_1,...,m_n)\in  M^+_{k-l}$, then the resulting $(m_1,...,m_n,\pm\frac{1}{2}l)$
form the entire set $M^+_k$. Also, for two different $l\not= l'$,
\begin{equation}
(m_1,...,m_n,\pm\frac{1}{2}l)\ \not=\ (m_1,...,m_n,\pm\frac{1}{2}l').
\end{equation}
This proves the following recurrence relation,
\begin{equation}
N^+_k\ =\ 2N^+_{k-1} + ... + 2N^+_0 + 1.\label{rec+}
\end{equation}
The (unique) solution is
\begin{equation}
N^+_k\ =\ 3^k.
\end{equation}
In conclusion,
\begin{equation}
N(a) \le 3^k\, .\label{N+}
\end{equation}

To find a lower bound for $N(a)$, we use the inequality
\begin{equation}
\sqrt{|m_i|(|m_i|+1)}\ \le\ |m_i|+\frac{1}{2},
\end{equation}
and consider the number $N^-_k$ equal to $1$ plus the number of
elements in the set
\begin{equation}
M^-_k\ :=\ \{ (m_1,...,m_n)\ \Big|\ n\in{\mathbb{N}},\
0\not=m_1,...,m_n\in\frac{1}{2}{\mathbb{Z}},\
\sum_{i=1}^n (|m_i|+\frac{1}{2})\ \le\ \frac{k}{2}\}\,
\end{equation}
Notice that this time, ignoring the constraint that the elements  $m_i$ of each
sequence  sum to zero (\ref{mzero}) makes an
in-equivalence  relation between $N(a)$ and $N^-_k$ a priori not known.
But let us postpone this problem for a moment. Using the same
construction as above, we find the recurrence relation satisfied by
$N^-_k$,
\begin{equation}
N^-_k\ =\ 2N^-_{k-2}+...+2N^-_0 + 1.\label{rec-}
\end{equation}
The unique solution is
\begin{equation}
N^-_k\ =\ \frac{2}{3}2^k + \frac{(-1)^k}{3}\, .\label{N-}
\end{equation}
A lower bound for $N(a)$ is  the number ${N'}^-_k$ of the elements of
$M^-_k$ which additionally satisfy $m_1+...+m_n=0$,
\begin{equation}
{N'}^-_k \ \le N(a).
\end{equation}
A statistical physics argument giving the value of the desired
number ${N'}^-_k$ is as follows (this argument is due to Meissner, who also
provided an exact proof in \cite{meis}). 
We can think of each sequence $(m_1,...,m_n)$ as of
a sequence of random steps on a line. The total length
of each path is bounded by $k$ owing to  the inequality in the definition
of the set  $M^-_k$. The number of sequences in
$M^-_k$ of a given, fixed value of the sum
\begin{equation}
m_1+...+m_n = \delta
\end{equation}
depends on $\delta$. The average value of $\delta$ is $\bar{\delta}=0$.
 For large values of $k$, the number of the paths corresponding to the random
walk distance $\delta$  should be given by the Gaussian function
$\frac{C}{\sqrt{k}} e^{-\frac{\delta^2}{\beta k}}\, \ \, N^-_k$.
In particular, the value
\begin{equation}
{N'}^-_k\ = \frac{C}{\sqrt{k}}N^-_k
\end{equation}
corresponds to $\delta=0$.

Summarizing,
\begin{equation}
\frac{C}{\sqrt{k}}N^-_k\ \le\ N(a)\le N^+_k,
\end{equation}
where the numbers $N^-_k$ and $N^+_k$ were calculated in (\ref{N+},\, \ref{N-}).
Therefore the entropy is bounded in the following way
\begin{equation}
\frac{\ln 2}{4\pi\gamma\lp^2}a + o({a}) \le  S(a) \le
\frac{\ln 3}{4\pi\gamma\lp^2}a. \label{entroest}
\end{equation}

We are in a position now, to compare our results with those of
\cite{abk}. The mistake  made in \cite{abk}  in the combinatorial
formulation (see (44) in that paper) is an assumption, that one
can order the $j_i$ labels in (\ref{phys}) such that $j_1\le ...
\le j_n$, and impose this as a constraint on `admissible' states
we count.  As a consequence, the number of the states contributing
to the entropy was underestimated: the analysis of \cite{abk}
implied that for large $a$  the leading contribution to the number
of sequences comes only from the sequences $(m_1,...,m_n)=(\pm
\frac{1}{2},...,\pm \frac{1}{2})$. The resulting number $S_{\rm
ABCK}(a)$ proposed for the value of the entropy was
\begin{equation}
S_{\rm ABCK}(a)\ =\ \frac{\ln 2}{4\pi\sqrt{3}\gamma\lp^2}a + O(a).
\label{entropyabk}
\end{equation}
The comparison with our (\ref{entroest}) shows that {\it the contribution from
the higher half integers $|m_i|>\frac{1}{2}$   can not be neglected.}

The next consequence of the estimate (\ref{entroest}) is a necessary condition for
the agreement of the  entropy $S(a)$ with the Bekenstein-Hawking
entropy  $a\,/\,4\lp^2$; the condition is that the value of the Barbero-Immirzi
parameter $\gamma$ is bounded in the following way,
\begin{equation}
\frac{\ln 2}{\pi}\ \le\ \gamma\ \le \frac{\ln 3}{\pi}.\label{gammabound}
\end{equation}
On the other hand, the value proposed for the Barbero-Immirzi parameter
by counting the sequences $(\pm 1/2,...,\pm 1/2)$ was
\begin{equation}
\gamma_{\rm ABCK} \ =\ \frac{\ln 2}{\pi\sqrt{3}}.\label{gammaabk1}
\end{equation}

In the accompanying paper \cite{meis} the recurrence relations
(\ref{rec-},\ref{rec+}) are generalized to an appropriate generalized
recurrence relation satisfied by the desired number $N(a)$ itself. 
Next, the combinatorial problem formulated in the previous section is
solved completely. The method is remarkably powerful. The result 
of \cite{meis} is
\begin{equation}
S(a)\ =\ \frac{\gamma_{\rm M}}{\gamma}\frac{a}{4\lp^2} - \frac{1}{2}\ln a + O(1), \label{me}
\end{equation}
where $\gamma_{\rm M}$ is a constant defined by the following
equation
\begin{equation}
\frac{1}{2}-\sum_{0\not=j\in\frac{1}{2}{\mathbb{N}}}
e^{-2\pi\gamma_{\rm M}\sqrt{j(j+1)}}=0.  \label{g}
\end{equation}
The numerical value of $\gamma_{\rm M}$ is calculated in
\cite{meis}.

An interesting historical remark  is (it has been unraveled to us
recently, after our results were derived)  that before coming to
their final result (\ref{entropyabk}, \ref{gammaabk1}), Ashtekar,
Baez and Krasnov had established certain estimate. The inequality
had been
\begin{equation}
S_{\rm bh}(a)\ \le\  \ \frac{\gamma_0}{\gamma}\frac{a}{4\lp^2} + O({a})
\end{equation}
where the constant $\gamma_0$ had been defined by the  same equation as 
as $\gamma_{\rm M}$ above. 
But since this result emerged as an upper bound only, finally it
was replaced by the improper one (\ref{entropyabk},\ref{gammaabk1}).
\bigskip

\section{The spin probability distribution}
Given any value $a$ of the classical horizon area, and the $N(a)$ quantum 
states of the horizon labeled by all the finite sequences $(m_1,...,m_n)$
of the non-zero half-integers such that (\ref{N(a)}) one can fix any arbitrary
\begin{equation}
0\not=m\in\frac{1}{2}{\mathbb{Z}}
\end{equation}
 and  consider the subset of states corresponding to the sequences such that 
\begin{equation}
m_1=m .
\end{equation}
Denote the number of the elements of this subset by $N_{(a)}(m)$.  
The ratio
\begin{equation}
P_{(a)}(m)\ :=\ \frac{N_{(a)}(m)}{N(a)}
\end{equation}
can be considered as the probability that the first puncture is labeled by
$m_1=m$. The question is, what $P_{(a)}(m)$ is when $a$ is large compared
to the minimal area $a_m$ created at the puncture\footnote{This issue has been raised recently by John Baez.}, 
\begin{equation}
a_m\ =\ 8\pi\gamma\lp^2\sqrt{|m|(|m|+1)}.
\end{equation}  
(Notice, that the answer could  make no probabilistic sense,
if the $\lim_{a\rightarrow \infty}P_{(a)}(m)$ were $0$ for all the values
of $m$, for example.)
The number $N_{(a)}(m)$ defined above can be also thought of as
a number of all the finite sequences $(m_2,...,m_n)$ such that 
 \begin{equation}
\sum_{i=2}^nm_i\ =\ -m,\ \ \ \ \sum_{i=2}^n\sqrt{|m_i|(|m_i|+1)}\ \le\
\frac{a-a_m}{8\pi\gamma\lp^2}.\label{Na(m)}
\end{equation}
By the same randon walk argument as the one used above in the treatment
of the number ${N'}^-_k$ (as before, see \cite{meis} for a proof) we have
\begin{equation}
N_{(a)}(m) \ =\ e^{-\frac{m^2}{a-a_m}}N(a-a_m) + ...   
\end{equation} 
where the other terms can be neglected when we  go to the limit  
$a\rightarrow\infty$. Since the number $N(a)$ is given by exponentiating  (\ref{me}),
we can see that in the limit $a\rightarrow \infty$
\begin{equation}
P_{(a)}(m)\ \rightarrow\ e^{-2\pi\gamma_{\rm M}\sqrt{|m|(|m|+1)}}\ =: P(m).
\end{equation}
Now, the equality (\ref{g}) means that the limits $P(m)$ of the probabilities
still sum to $1$,
\begin{equation}
\sum_{0\not=m\in\frac{1}{2}{\mathbb{Z}}}P(m)\ =\ 1.
\end{equation}

That (limit) probability distribution $P(m)$ can also be
written in terms of the entropy $S(a_m)$ corresponding to
the area $a_m$, namely
\begin{equation}
P(m)\ =\  e^{-S(a_m)},
\end{equation}
regardless of whether we fix the value of $\gamma$ by  
the agreement with the Bekenstein-Hawking entropy or not. 
\bigskip

{\bf Acknowledgments} We benefited a lot from the discussions with
Abhay Ashtekar, John Baez, Alejandro Corichi, Kirill Krasnov
and  Marek Napi\'orkowski.
Krzysztof Meissner, the referee of the diploma thesis of one us (MD), 
helped us to understand the relevance and irrelevance
of the  constraint on the quantum excitations of the C-S field.
JL was  partially supported by Polish Committee for Scientific Research
(KBN) under grant number 2P03B 12724.


\begin{thebibliography}{99}

% 1
\bibitem{thiembook} Thiemann T 2003 Lectures on Loop Quantum Gravity in{\it Quantum Gravity: From Theory to Experimental Search} Lecture Notes in Physics {\it Preprint} gr-qc0112038\\
Thiemann T 2004 {\it Modern Canonical Quantum General Relativity} (Cambridge: Cambridge University Press) at press gr-qc/0110034

%2
\bibitem{rovbook} Rovelli C 1998 Loop quantum gravity {\it Living Rev. Rel} {\bf 1} 1

%3
\bibitem{alrev} Ashtekar A and Lewandowski J 2004 Background Independent
Quantum Gravity:
A Status Report {\it Class. Quantum Grav.} {\bf 21} R1-R100 gr-qc/0404018


\bibitem{krasentropy} Krasnov K 1998 On Quantum Statistical Mechanics of a Schwarzschild Black Hole {\it Gen.Rel.Grav.} {\bf  30} 53-68

%7
\bibitem{roventropy} Rovelli C 1996 Black Hole Entropy from Loop Quantum Gravity {\it Phys.Rev.Lett.} {\bf 77} 3288-3291


%4
\bibitem{rsarea} Rovelli C and Smolin L 1995 Discretness of area and volume in quantum gravity {\it Nucl. Phys.} {\bf B442} 593-622; Erratum: {\it Nucl. Phys. } {\bf B456} 753

%5
\bibitem{alarea} Ashtekar A and Lewandowski J 1997 Quantum theory of geometry I: Area operators {\it Class. Quant. Grav.} {\bf 14} A55-A81

%6

%8

\bibitem{abk} Ashtekar A, Corichi A and Krasnov K 1999 Isolated Horizons: The
Classical
Phase Space {\it Adv. Theo. Math. Phys.} {\bf 3} 419\\
Ashtekar A, Baez J C, Krasnov K 2000 Quantum geometry of isolated horizons and black hole
entropy {\it Adv. Theo. Math. Phys.} {\bf 4} 1-95


%9
\bibitem{bekhaw} Bekenstein J D 1973 Black holes entropy {\it Phys. Rev.} {\bf D7} 2333\\
Bekenstein J D 1974 Generalized second law of termodynamics in black hole physics {\it Phys. Rev.} {\bf D9} 3292\\
Bardeen J W, Carter B and Hawking S W 1973 The four laws of black hole mechanics {\it Commun. Math. Phys.} {\bf 31} 161

\bibitem{Dre} Dreyer O 2003   Quasinormal Modes, the Area Spectrum, and Black Hole Entropy  {\it Phys.Rev.Lett.} {\bf 90}  081301


%12
\bibitem{meis} Meissner KA 2004 Black Hole etropy in Loop Quantum Gravity
gr-qc/0407052

%13
\bibitem{adlm} Ashtekar A, Domagala  M, Lewandowski J, Meissner KA 2004 in preparation

\bibitem{IH} Ashtekar A and Krishnan B 2004 Isolated and Dynamical Horizons and Their
Applications gr-qc/0407042 \\
Ashtekar A, Beetle C, Dreyer O,  Fairhurst S,  Krishnan B, Lewandowski J,
 Wisniewski J 2000  Generic Isolated Horizons and their Applications
 {\it Phys.Rev.Lett.} {\bf 85}  3564-3567

%14
\bibitem{pinek} Domagala M 2004 Kwantowy operator pola powierzchni oraz
entropia czarnej dziury {\it Diploma Thesis, Uniwersytet Warszawski, Warsaw,
Poland}


% Praca o klasycznej czarnej dziurze -------------------------------------------
\end{thebibliography}
\end{document}